\documentstyle[12pt,twoside,fleqn,espcrc1,epsfig]{article}

\def\be{\begin{equation}}
\def\ee{\end{equation}}
\def\bea{\begin{eqnarray}}
\def\eea{\end{eqnarray}}  

\title{Collective flow and QCD phase transition}       

\author{H.~Sorge
       \address{Department of Physics \& Astronomy \\
        SUNY at Stony Brook, Stony Brook, NY 11794, USA}
        \thanks{E-mail: Heinz.Sorge@sunysb.edu}}

\begin{document}
\maketitle

\begin{abstract}
In the first part I discuss the sensitivity of collective matter expansion
in ultrarelativistic heavy-ion collisions to the transition between quark and
hadronic matter (physics of the softest point of the Equation of State).
A kink in the centrality dependence of elliptic flow has been suggested as a 
signature for the  phase transition in hot QCD matter. Indeed, preliminary 
data of NA49 presented at this conference show first indications for the 
predicted kink. In the second part  I have a look at the present theories
of heavy-ion reactions. These remarks may also be seen as a critical comment 
to B.~Mueller's  summary talk  (nucl-th/9906029)  presented at this conference. 
\end{abstract}

\section{A Kink in the Centrality Dependence of Elliptic Flow}

Information
about the quark-gluon plasma (QGP)
 and the phase transition region has become available
with the advent of  powerful
lattice  simulations  of quantum chromodynamics (QCD). 
Most notably, it has been
shown that chiral symmetry is restored at rather low temperatures
(in the range 140 to 170 MeV). Furthermore, the Equation of State (EOS)
varies rather rapidly in the transition region. It is not clear yet
whether the transition is of weak first-order  or  just a rapid
cross-over between the two phases.
The EOS extracted from the lattice clearly displays the transition
from hadron to quark-gluon degrees of freedoms.
Pressure and energy density approach the ideal
Stefan-Boltzmann values at temperatures $\ge 3 T_c$.
A generic feature of the EOS in the transition  region is the
presence of the so-called ``softest point of the EOS''
\cite{MHS95,RPM95}
related to the effect that the energy density may jump with
increasing temperature but the pressure  does not.

The collective transverse flow which
develops in the heavy-ion collisions reflects on  the properties
of the EOS.
Usually, one distinguishes various types of transverse flow,
the radial (isotropic component), directed (sideward kick
in the reaction plane) and the elliptic flow, the latter being
a preferential emission either along the impact parameter axis
or out of the reaction plane (squeeze-out) \cite{PV98}.
The general idea why a phase transition may show up in flow observables
is rather straightforward.
At densities around the softest point the
 force driving the matter expansion gets weakened.
A long time ago, van Hove has suggested that the multiplicity
dependence of  average transverse momenta
may display a plateau and a second rise \cite{vH82} which was not seen, 
however.
Some time ago I have suggested that
the elliptic  flow may be 
better-suited  to identify a first-order type phase transition
\cite{HSplb97}. 

Elliptic flow in the central region of ultrarelativistic
collisions is driven by the almond-shape of the participant matter
in the transverse plane \cite{ollitrault}.
It was argued in \cite{HSplb97} that elliptic flow may be more
sensitive to the early pressure than the isotropic radial flow.
``Early'' and ``late'' is defined by the time scale set by
the initial transverse size
 $r_t=\sqrt{\langle x^2 +y^2 \rangle}$
 of the reaction region.
The  time dependence of flow build-up has been studied first using the 
transport  model RQMD, for radial flow in \cite{vHSX98}  and for
 elliptic asymmetries in \cite{HSplb97,HSprl99}. 
From these results it has been inferred that radial flow continues
to develop  for much longer time than its azymuthal asymmetry. 
\begin{center}
\begin{figure}[h]
\centerline{\epsfig{figure=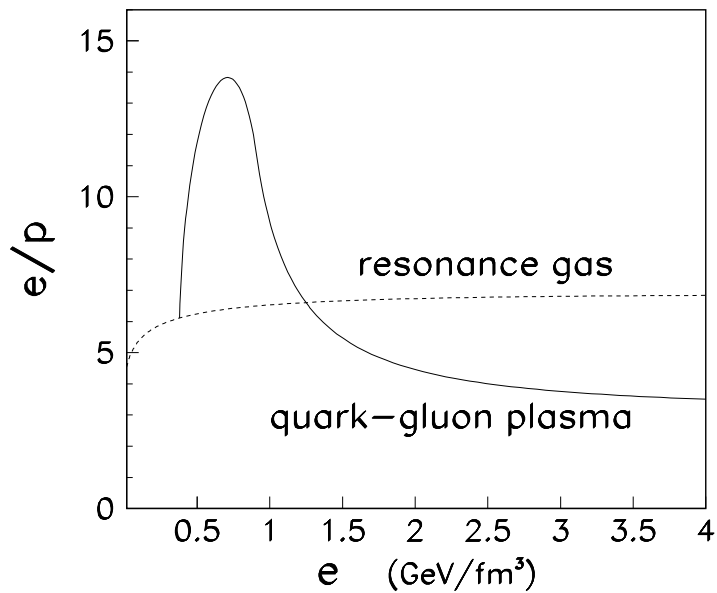,width=8.0cm}}
\vspace{-1.0cm}
\caption{
 Equation of states implemented into RQMD:
 ratio of energy density $e$ divided by pressure $p$.
 The dashed line  represents the resonance gas EOS,
 the solid line  the EOS including a first order
  phase transition
    with $T_c$=160 MeV.
 }
\end{figure}
\end{center}
 
One reason for the larger sensitivity of the elliptic flow to early
pressure may be that  the generated flow asymmetry  works against
its cause and diminishes the spatial asymmetry on a time scale proportional to
  $\sqrt{\langle y^2 \rangle}-\sqrt{\langle x^2 \rangle}$.
 U.~Heinz and Collaborators have  studied  recently   elliptic flow 
 in a hydrodynamic  framework \cite{PSH99}. 
 Their  important finding is
 that the net effect of the phase transition is much less than naively
 expected based on this argument. 
  The effect due to the smaller pressure near the phase transition
 is largely canceled. The system also spends more time in the transition 
 region.
On the other side, the {\em non-ideal } character of the expansion
 dynamics may  be the leading cause that development of elliptic flow is
 shut off earlier than of radial flow. 
The elliptic asymmetry  is proportional to the difference  
 between the flow strengths in  $x$
 (parallel to impact parameter) and $y$ direction. Thus it is
 more fragile than radial flow.  Partially free motion
 (viscosity in hydrodynamic language)
 tends to wash out the pressure-driven asymmetries.
 Obviously, these effects will be  more pronounced in the later
 dilute stages of the reaction.        
   It is amuzing that  non-ideal effects from viscosity in the
   low-density stage
   may be helpful to infer information about the pressure in the
   high-density region.          

Recently, I  presented 
a novel signature of the QCD phase transition based on elliptic flow. 
A rather characteristic centrality dependence of the elliptic flow
-- a ``kink'' -- is predicted
if the created system passes through the softest region of the EOS in the
heavy-ion reactions \cite{HSprl99}.
 Heiselberg and Levy have also put forward  arguments why a   
1st order transition may  be reflected in the $b$ dependence of elliptic
flow \cite{HHL98}.

  Fig.~1 displays the resonance-matter EOS based on  counting  the
  propagating quasi-particles in RQMD. In addition, an  EOS
  is shown which 
   is calculated  from a
  quasi-particle model  of quarks and gluons with dynamical thermal
  masses  \cite{P96}.
  I have chosen this EOS, because it provides a good fit to lattice data.
  The EOS contains a 1st order transition at $T_c$=160MeV with a latent heat
   of 467 MeV/fm$^3$.
   In the following results of calculations  with the  RQMD model 
  which includes either one of these two EOSs will
  be  compared. 

 Let me first shortly describe how the 
  EOS with phase transition is implemented
 into the RQMD model\cite{SOR95}. 
   In RQMD
  nucleus-nucleus collisions are calculated in a Monte Carlo type fashion.
  While the nucleons from each of the colliding nuclei pass through each other,
   they are decomposed into constituent quarks.
  Strings may be excited, and overlapping strings fuse
   into ropes (with larger chromoelectric field strength).
    After their decay and fragmentation  secondaries emerge and may interact
   with each other. Formed resonances are treated as unstable
   quasi-particles.  This  leads to a resonance gas    EOS
   if there are no corrections  from other interactions.
  The QCD dynamics in the phase transition region is not well understood.
   Even if there is a quasi-particle description it is not obvious
    which one  of the  possible choices (strings, constituent quarks,
    partons, either  massless or with dynamical masses)
    is to be prefered.    In this situation  I 
   stick to the implemented degrees of freedom and modify the collision
   term instead. Since  hydrodynamics is  expected to be a reasonable
    approach  for the transverse dynamics in the ultradense stage,
    the EOS should be the most relevant ingredient
     for the expansion dynamics anyway.
  It is well-known that different treatment of
   interactions between quasi-particles may modify the EOS \cite{DP96}. 
  The standard collision term in  RQMD  is manufactured such
   that it does not contribute to the pressure. Now, we
    let each quasi-particle interact elastically with a neighbor
   after any of the standard collisions such that the average 
   momentum change leads to a prescribed change of the
  total pressure according to the EOS $P(e)$.
   $e$ is the energy density. One should note that this is a reduced
   EOS, because the temperature $T$ is eliminated. This way we can
   use the pressure modification not just in equilibrium but also
   in the non-equilibrium situation. Furthermore, the EOS enters into
   hydrodynamic equations via $P(e)$. Here we are mainly interested in
   studying the relation between hydrodynamic flow and EOS.

Let me now turn to a discussion of how a 1st order phase transition affects
the centrality dependence of elliptic flow. Experimentally, the elliptic flow
can  be extracted from 
the  azymuthal asymmetry of final hadrons
\be
  v_2= \langle  \cos (2\phi )  \rangle
\ee
as a function of centrality. 
Of course, the spatial asymmetry of the reaction zone which is
correlated with the asymmetry of the participant nucleons in the ingoing nuclei
\be
     \epsilon=
          \left(
             \langle  y^2  \rangle
            -  \langle  x^2  \rangle
          \right)/
          \left(
             \langle  x^2  \rangle
            +  \langle  y^2  \rangle
          \right)
\ee
is itself a function of the impact parameter. Trivially,  $v_2$  goes
to zero for  very small and very large impact parameters because of
 the corresponding behaviour of the  spatial asymmetry.
We may disentangle the effects from geometry and
dynamics
by defining the  scaled flow asymmetry as
\be
 A_2 = v_2 /  \epsilon
   \quad .
\ee
 $A_2$ represents the dynamical response of the created system to the
initial spatial asymmetry.
\begin{center}
\begin{figure}[h]
\centerline{\epsfig{figure=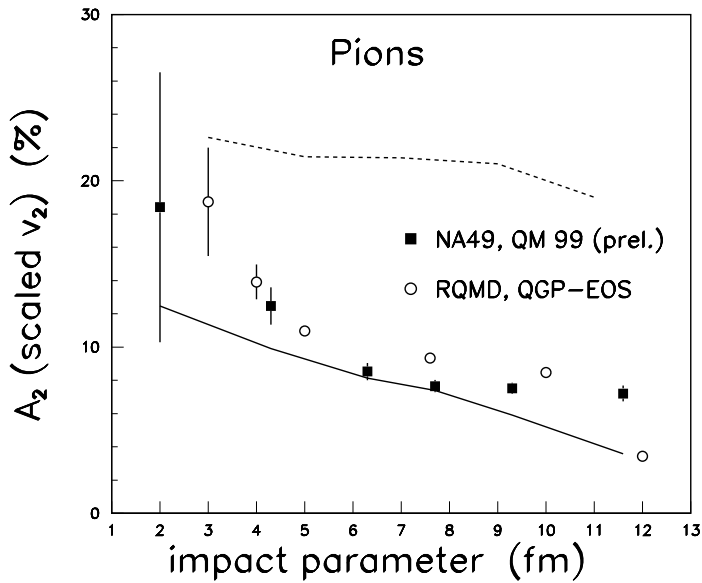,width=8.0cm}}
\vspace{-1.0cm}
\caption{
 Centrality dependence of  scaled flow asymmetry $A_2=v_2/\epsilon$  for pions 
 in Pb(158AGeV) on Pb: comparison of RQMD results
  employing resonance gas EOS (solid line) versus
  EOS with phase transition (open dots),
  Hydrodynamics (dashed line). Preliminary NA49 data presented
   by A.~Poskanzer at this conference are solid squares.  
 }
\end{figure}
\end{center}

\vspace{-1.5cm}
We display the scaled flow asymmetry $v_2/\epsilon$  for pions 
in the reaction Pb(158AGeV) on Pb
versus impact parameter $b$ in Fig.~2.
Extracted   values from
hydrodynamic  calculations  
\cite{ollitrault}  show  essentially
no centrality dependence, except for the grazing collisions.
All information on the EOS is essentially integrated into one number.
For realistic EOSs this number varies between 0.16  \cite{PSH99}
and 0.24 \cite{DT99} for pions. It  depends only weakly on other input like
the initial or freeze-out temperature. 
We show a typical hydrodynamical result \cite{DT99}  in Fig.~2.
The centrality dependence from hydrodynamics is in marked contrast to the
transport calculation which includes the  non-equilibrium aspects
of the dynamics.
Without phase transition, the asymmetry factor   $v_2/\epsilon$ calculated from
RQMD  simply increases monotoneously
with centrality -- approximately linearly with the initial system size in the
reaction plane ($\sim 2 R_{Pb}-b$).
Partial thermalization
 -- so-called pre-equilibrium softness \cite{HSplb97} -- 
initially, viscosity
 and system-size dependent freeze-out  play the major role here.
 Since the  resonance-gas EOS $P \sim e$  has no structure, 
  no structure develops in the $b$ dependence of $v_2/\epsilon$ either. 
This monotoneous dependence  of  the RQMD result without phase transition
 is also displayed in Fig.~2. (The author is indebted to A.~Poskanzer
 and Collaborators at LBNL who have actually done these RQMD
 calculations and provided the result.)
  The RQMD calculation with the EOS with phase transition displays
  a much more interesting centrality dependence of the scaled
 flow asymmetry $v_2/\epsilon$. In not too peripheral collisions the
  system spends more and more time in the ``mixed phase'' where
  pressure is low. Thus the increasing  initial energy density, 
    reaction time  and size
 which are all helpful to develop the asymmetry 
 are counteracted  by the increasing softness   of the matter
   around the softest point of the EOS. 
  A plateau-like
   $b$ dependence of $v_2/\epsilon$ develops (for moderate latent heat values 
   400--500MeV/fm$^3$ as in this particular calculation). 
  At some centrality, the softening from mixed phase
   has developed full strength ($b \approx 5$ fm in the calculation).
  Further increase of the centrality means that initial high
   pressure stage from the QGP is added. Therefore $v_2/\epsilon$
  starts to rise again strongly for $b< 5$ fm. 
  This kink in the centrality dependence of scaled elliptic flow is
  a combined effect of softest point in the EOS and non-equilibrium
   due to smallness of the  system. 
  Neither one of these effects separately produces
   the characteristic ``kinky'' $b$ dependence. This has been
  demonstrated by
   the hydrodynamical with phase transition
   and the RQMD calculation without.
   It should also be noted that the RQMD calculation approaches closely the
    hydrodynamical result for very central collisions. This is
    precisely what one would naively expect about the system size
    dependence of non-equilibrium corrections. 
   
  At this conference, A.~Poskanzer has presented first experimental
  data for the scaled flow asymmetry $v_2/\epsilon$  (see his contribution to
  these Proceedings).
  The RQMD prediction was taken from \cite{HSprl99} and has narrower
  rapidity coverage than the preliminary NA49 data. A quantitative
  comparison has therefore to be taken with some grain of salt.
   Nevertheless, it appears that the experimental measurement follows
   more closely the trend of the result if the phase transition 
   is included in the EOS. Of course, in view of this exciting
   indication one would like to see data  (and model results)
   with much better statistics
   in the region $b< 5$ fm.
    $p_t$ cuts  (low $p_t$ pions  show an even more pronounced kink)
    and consideration of other hadrons (protons, strange particles)
   would also be of interest. 
 
  Study of the $b$ dependence may also be very useful for Au on Au
  collisions at RHIC. What to expect depends essentially on the
   (so far unknown) particle densities. If the RQMD approach continues
   to describe hadron multiplicities well, such a kink 
   can  also be expected.
   If initial particle production is much higher than RQMD predictions,
   the hydrodynamic result may be closer to reality. 
    Elliptic flow in 
    collisions at RHIC has been studied using a parton cascade
     model \cite{ZGK99}. 
   Some dependence on viscosity (magnitude of cross section)
   has been found, but the hydrodynamic result 
    (for the hard EOS without phase transition $P=e/3$) 
    has been qualitatively
     confirmed. 
    Elliptic flow holds much promise to clarify the structure 
    of the EOS in the phase transition region -- at finite
    temperature and also baryon density (see P.~Danielewicz's talk). 

\section{Quo Vadis, Heavy-Ion Theory?}

Recently, B.~Mueller has provided the write-up of his summary talk
given at QM '99 \cite{bum99}. 
He expresses his opinion  what should be done in
heavy-ion theory and -- more explicitly -- what should not be done. 
(According to him the study with RQMD whose results have been
  presented in the first section should {\bf not} have been done.)
Taking his contribution at its most positive aspect,  it 
expresses a need  
 to reflect  on the direction of heavy-ion physics.
  I   feel this need as well, not just
in view of RHIC coming up soon but also due to the uncertainty
about the merits and prospects of heavy-ion physics. 
On the other side, I strongly disagree with the direction for heavy-ion
theory which  Mueller advocates. 
The problem is not mainly with this
particular direction but its ``absoluteness'', i.e. that it is just one
direction out of several possibilities. 
 Expression of
beliefs and dogmas are no substitute for proofs or disprovals
in science. 
Mueller's feverish rhetorics (``nonsense'') or threats
(``do not even think of applying these models at RHIC'') would be expected
in preachings of a sect but not in a summary talk at a physics
conference \cite{bmtalk}.

 There are  three types of approaches in
 theoretical heavy-ion  physics , the  thermal fireball, hydrodynamics
  and transport theory. 
 These
 different approaches complement each other and
  will continue to persist side-by-side.            
  Their mutual usefulness is not just an abstract statement 
  but has been proven
  in the past and at this conference as well. 
  Indeed, a five-minute analysis of transverse hadron spectra
   in the framework of the static thermal fireball model 
   gives reasonable values for kinetic freeze-out temperature 
   (100--140 MeV)
   and flow velocity
   ($v = 0.4-0.6 c$) for Pb(158AGeV) on Pb collisions.  
   Furthermore, particle ratios tell that chemistry
   is different and points to a clearly higher freeze-out ``temperature''. 
   Fine. 
   (It has been predicted by RQMD
   at a time when Stachel, Braun-Munzinger and Collaborators were
   arguing for a common kinetic and chemical freeze-out based on
    fireball results for single-particle spectra \cite{BMS96}.)
  Do we need more  sophistication than provided by the fireball model? Yes. 
If not for anything else, different independent approaches 
may be used to  uncover elementary
mistakes in too simplistic models. 
Mueller's talk provides a good example how reliance on 
the most simple one -- the static fireball -- 
produces disastrous errors. 
Mueller applauds Wiedemann to have presented 
(at this conference) 
``a highly consistent
picture (as) emerging from these measurements. The single particle
momentum spectra, the $p_t$-dependence of the pair correlations,
and the fragment yields all can be explained by a freeze-out
from a thermal, dilute hadronic fireball with a final temperature
(at the SPS) around $T_f\approx $ 120 MeV, an average transverse flow
velocity $\langle \beta _f \rangle \approx $ 0.55,  and a baryon chemical
potential $\mu _f\approx $ 250 MeV. The rms radius of the fireball
at freeze-out is slightly above 10 fm, and the average freeze-out time
is about 6 fm/c  spread over a rather short period of 3 fm/c''.
These numbers refer to the system Pb(158AGeV) on Pb. 

Now, $s= s_0 + \beta \cdot t $  (constant velocity) or
 $s= s_0 + \frac{1}{2} \beta \cdot t $ 
   (constant  acceleration) 
 are indeed  formulas which somebody
talking about velocities should be aware of.  Let us take the most extreme
case in which all flow velocity
 is created instantaneously (quite unrealistic), the first formula.
 How  can a system of initial rms size 4 fm 
  (given by the wounded nucleon participants of the colliding two
   Pb nuclei)
 expand to  10 fm within just 6 fm/c -- but with half the velocity of light?
 A more realistic
  acceleration history, e.g. as given by the second formula,
    magnifies the problem.
 R.~Stock presents somewhat different numbers, based on the same fireball
 approach and for the same system \cite{RS99}.
  However, his -- slightly higher -- value for the ``life-time''  (8 fm/c)
  but  same value for transverse expansion as Wiedemann's (factor 2.5)
 appears also incompatible with any  dynamical  expansion model, be it
 hydrodynamic or transport approach. 

This example may  clarify to which erroneous conclusions  a 
  self-imposed restriction to the simplistic fireball model 
  which has not much to say about time evolution (dynamics) 
  may lead to.     
 A look at this author's  
   prediction for final flow velocities, 
  (average) freeze-out ``temperatures'', source sizes  and life times
  based on transport calculation with RQMD  \cite{HSplb96}
    might have been helpful to avoid such hazard. 
 I should add for non-initiated readers
  that the value of these numbers for life-time etc. 
  is of tremendous importance
  for a  qualitative  understanding of the expansion dynamics. Is it
   an ``explosive expansion pattern'' as Stock put it  (a bang)
   or soft transverse expansion 
    which I among others  \cite{Dum97,MHS98}  have argued for, 
  a ``fizzle''  (a word which I borrow from E.~Shuryak)?

The content of the debate with  Mueller  is 
which strategy may be successful
to understand and approximate quantum chromodynamics (QCD)  of dense
matter as it is produced in ultrarelativistic  heavy-ion reactions. 
If we forget non-equilibrium features, it is the physics of QCD
at 1--2 $T_c$, with the ``critical temperature'' around 150--170 MeV.
Since equilibrium situation is much simpler, I start with some remarks 
about finite $T$ QCD. 
Smilga among others has pointed out that the physics of the state
above and close to $T_c$ is a theoretical ``no man's land'' \cite{Smi99}. 
I share this opinion. 
Neither
perturbative QCD nor interactions between (quasi-)hadronic
states appear justified for a description. 
At hadron densities of one per cubic fermi or equivalent quark densities
even the pragmatic question  from which single-particle base  to start
(hadrons or quarks)  is not easily answered. 
We have to do the physics 
 at RHIC and LHC 
  facing the fact
that there is a wide hole in our knowledge of QCD in the most interesting 
temperature region. 

The gap in our knowledge cannot be filled by statements of belief like
the one from Mueller's talk:
``the picture of QGP as a plasma of weakly interacting quasi-particles
 \ldots may work until very close to $T_c$''. Maybe, but maybe not.
I do not want to be mistaken. The quasi-particle approach
is a very important strategy to resolve some of the problems
which have plagued perturbative calculations for thermal QCD so far. 
In fact, to my knowledge I have been the first person to utilize 
the EOS as calculated within such a quasi-particle  approach for 
(effective) quarks
and gluons in  calculations of heavy-ion reactions. 
On the other side, nobody knows yet how good the quasi-particle picture is,
 in particular in the range 1--2~$T_c$. 
 Agreement on the EOS level does not tell much, because many rather
 different interacting systems may have the same EOS. 
 A ``screening mass'' is a much better measure of correlations
  in the plasma than the EOS. 
 Kajantie et al. have erected a strong warning sign in that respect.
 This group has shown that the
 perturbatively calculated screening mass approaches the real one
 only at astronomically large temperatures (at which QCD itself is invalid)
  \cite{Kaj97}.
 Roughly, in the temperature range of  interest the real and
 the  perturbative  screening mass differ by a factor of 3--4. That means
  that the perturbatively calculated  viscosity 
  (among other transport coefficients)
  can be expected to be
  wrong by the square of that number, i.e. an order of magnitude.

Of course, non-equilibrium situation in  heavy-ion reactions is much
more complicated than the finite-temperature case. 
Therefore all what was said before about the lack of knowledge
applies even  more in this situation. Furthermore, for 
heavy-ion reactions we would need to know how a system of quarks and
gluons hadronizes. Since that is connected to the mystery of 
(de-)confinement a resolution belongs yet to the realm of Utopia.
Either one gives up in such a situation or one resorts to 
{\em modeling} which is the typical strategy in physics.
As usual, there are  flawed, 
bad and  better models, and we will have to say
more about this specifically lateron. 
Moreover, since no model can be completely trusted, it is very 
important to test a broad variety of models as much as possible
against the ``true'' theory and available data to sort out
generic features from arbitrary model details. 

In contrast, 
Mueller  argues against use of what he calls ``hadronic
models'' to describe  collisions among heavy nuclei at the SPS. According to 
 him  they are (partially) 
 pure fantasy, consequence of the imagination of its author. 
 Perhaps. Every model carries this risk. Unfortunately, Mueller did not
 reflect on that this might apply to his own model creation
 (together with K.~Geiger), the  parton cascade model as well.
 (The lattice data for the EOS 
  are in clear     
 conflict with a weakly coupled gas of almost massless partons.)
Actually, I know only of few hadronic models 
 applied at SPS energies,  
 Kahana$^2$'s  Lucifer, Kapusta's and Jeon's   Lexus 
and Cassing's  HSD.
 Mueller   subsumes ``string'' models like   RQMD 
  and   VENUS  into this category
  which I believe is a misspecification. 
  In these models in-going hadrons are destroyed  into multiple components
  on the valence and sea quark level. 
  Of course, they behave differently (create strings) than 
  perturbatively interacting partons. However, 
   in view of the theoretically unsettled situation it is not obvious
  a priori whether this should count positively or negatively. 
 
What are the pro's and con's  of the various types of 
 microscopic dynamical models?
 In order to discuss the range of applicability one should
  take into account that nuclear reactions are characterized by vastly
 different stages, schematically
\begin{itemize}
  \item
     {\bf the first fm/c}:
    destruction of the ingoing hadrons (nuclei) and
    multi-particle production,
   \item
   the {\bf hot, ultra-dense  stage }
    (which includes the possible phase transition), 
    and
   \item
   the {\bf dilute hadronic stage} until freeze-out.
   (The proper treatment of this stage is not contentious.)
\end{itemize}

 Main contenders for the ``initial-stage'' physics are
 ``soft production'' models based on Regge theory supplemented with the
 string picture   and ``semi-hard  production'' models
 (Mini-jets as studied by X.N.~Wang, M.~Gyulassy and K.~Eskola, Geiger's and
 Mueller's parton cascade and  at ultrahigh energies
 the semi-classical Weizs\"acker-Williams type approach advocated by
  L.~McLerran, R.~Venugopalan  and Collaborators).
 To some degree, the two pictures are complementary -- with strengths and
 weaknesses in different areas. 

  {\bf Perturbative QCD based approaches}: the natural starting point
  is to look at high $p_t$ particles, understand their properties
   and then march towards lower $p_t$. At which $p_t$ -- perhaps dependent
  on hadron species -- 
   does the model break down and ``soft physics'' (flow, etc.) sets in?
   Recently, X.N.~Wang  has studied  high $p_t$ pions at SPS energies
   (see his talk).
   Contrary to expectations about strong energy loss of partons in
   the dense medium 
    his finding
   is  that the partons seem to experience essentially no loss at all. 
   That is one of the most important results in HI physics of the past
   years.  Geiger's and  Mueller's parton cascade model (VNI) overshoots
    the WA93 data by a factor of 5--10 at high $p_t$ 
     \cite{SG98}
      where the model
   should work. Why not? Maybe  Mueller, Srivastava 
   and Collaborators could look for
   answers. My suggestion would be that the problem is in the
    space-time ordering of partons with different rapidity.
    A further hint of problems with present  parton cascade VNI
    has been provided by Longacre recently \cite{Ron99}. 
    Energy and momentum is created outside the future light cone
    of the two colliding nuclei. It 
    appears to be a strong violation of causality. 
    Recently, VNI has been tested by the OPAL group for
    W$^+$W$^- \rightarrow q\bar{q}'  q\bar{q}'/l\bar{\nu}_l $,
    a system of few hard partons  \cite{OPAL99}.  In contrast to
    all other tested pQCD based codes (PYTHIA, HERWIG, ARIADNE) 
    it failed badly in reproducing the event properties. 
   Structure  functions  are distributions in momentum space, but a
   parton cascade needs initial condition in phase space (space-time
   and momentum). 
   We urgently need more studies how to initialize a parton cascade
   in phase space. 

   {\bf 
  Regge theory, strings,
    baryon junctions, ropes and all that: cumbersome things of the past
   and irrelevant for RHIC physics?  }
   The ``excommunication'' of such non-perturbative phenomena from
   HI physics  is suggested in B.~Mueller's summary talk. 
   I beg to disagree. 
  There were always strong indications
  that QCD at large distances may be expressed as a theory of
   interacting strings. t'Hooft made the argument based on large
   $N_c$ in the 70s. Of course, the real reason 
    predates QCD. It was Regge theory  which was abstracted ``emprically''
   from  strong interactions in the soft domain.
    Everybody knows since Veneziano's work in '69 that
    string theory  is the natural candidate  underlying Regge physics. 
  It was one of the most important developments 
    initiated by Witten \cite{Wit98} and others   
   in the course of 
   last year (and has gone unnoticed  at  QM '99) that the
   idea of {\em duality } between gauge theory in the large $N_c$ limit
   and string theory may be put on firm grounds by extending 
  Maldacena's conjecture into the non-supersymmetric domain.
 
  There are many aspects of the string-gauge theory duality which may
   be fruitful to heavy-ion physics. One of them goes to the heart of
  the nature of the QCD phase transition: if gauge theory can be mapped
   to a certain string theory, then  string theories at finite $T$ 
  should not contain only the possibility of a limiting 
  (Hagedorn) temperature   but also of a phase transition.       
    It is noteworthy that the ``empirical'' QCD transition temperature
   extracted from the lattice ($\approx $ 160 MeV) coincides with the
   Hagedorn temperature  estimated from the spectrum of hadronic
  resonances. 
 
  Another important issue for RHIC physics is baryon stopping.
   In a string model like RQMD baryon shift is  a deeply non-perturbative
    process, the stripping of valence quarks off the ``baryon junction''
    (which is connected to the valence quarks via strings).
    D.J.~Gross, one of the fathers of QCD,  and Ooguri
      co-wrote a paper
     in which the large $N_c$ baryon wave function is constructed from
      $N_c$ strings connected via a junction using some
     super-gravity  solution (and employing the recently conjectured
    duality to gauge and string theory) \cite{DGO98}.
   A highly non-trivial prediction of the junction dynamics in RQMD is that
    baryon number is stopped more than the valence quarks.      
   Why not look for it experimentally in order to see whether it is
    a fictitious component of the model?
   
  Color ropes are flux tubes of chromoelectric field created by charges
  higher than the elementary $SU_N$ color charges. 
   In RQMD they form if strings would overlap. 
   (The idea of coherent superposition of gluon fields from
    valence quarks of different nucleons has also been the starting point of
    McLerran and Collaborator's work, however, in the perturbative domain.)
   In contrast to
  Mueller's claims, ropes have been studied from first principles,
   in lattice gauge theories  \cite{BER82}.
   Important properties needed for phenomenology like the transverse
   size independence  on the  color field strength 
   have been confirmed on the lattice  \cite{TRO93}.
   Whether they decay via Schwinger-type particle creation like it is
   assumed in RQMD I do not know. It may be tested experimentally,
   however. Coherent fields are {\em stronger } than incoherent ones.
   Therefore they are screened earlier.
   This effect is at the root of the result why  particle production
   at RHIC is so low in RQMD. A charged hadron dN/dy of approximately
   700  is predicted for central Au on Au collisions. Incoherent
   multiple-scattering models (independent whether they follow 
   parton cascade receipes
   or  Regge theory) predict a much higher
  rapidity density, 1500--2000. In a couple of months from now we will know
   the real answer and can review the existing theoretical attempts in light
    of the experimental findings.
   
   In his summary, Mueller provides one ``dramatic revelation''
   why strings should not be considered at SPS energies. It is a recent
    result by Bass et al. \cite{BAS98} according to which
    most of the energy density up to a time of 8~fm/c (in Pb+Pb)
    resides in sterile (=non-interacting) strings. 
   I find this result rather cumbersome, and it is not found in a
   string model  like RQMD. Generally, time scale of string fragmentation
    is governed by string tension and quark masses which 
   leads to the famous formation time of around 1 fm/c.
    Such formation time has been  observed
    in p+A experiments. 
    Which physics
     leads to a scale of 8 fm/c for string decay?
   (A possibility would be that a Lorentz factor $\gamma $ has sneaked 
    somewhere into the calculation.)
    
   Finally, I would like to add one comment concerning the
    modeling of the 
   the hot, ultra-dense  stage  of HI collisions 
     including the  phase transition. 
    Certainly one should explore the transport theory  of
     quark and gluon quasi-particles in the plasma
   and for   non-equilibrium situations. 
   Hopefully these concepts will work at temperatures not too far from 
    $T_c$, say 2 $T_c$. There are plenty of observables 
    in HI collisions at RHIC (and LHC) for which
    predictions are needed, photons and intermediate-mass dileptons,
     jet quenching etc.  -- with sensitivity to high-temperature physics. 
   Nevertheless, physics of the  phase transition will be quite a
   different situation. A factor of 2 on the temperature scale corresponds
    to a factor 16 or so on the energy density scale.
   There are two related -- practical -- reasons why use of                
    hadronic  quasi-particles is sensible, above $T_c$  and keeping all the
    caveats about unknown properties of these objects in mind.
    They come into play 
    why  I do not share Mueller's belief that
     coupling of parton cascade with hadronic ``afterburner''
    -- like being done by Bass et al. \cite{BAS99} -- is the only 
      reasonable strategy for transport calculations at SPS and at RHIC.
   One of the most important reasons to do 
   transport is to study the EOS in the phase transition region
    including non-equilibrium effects. 
   Seen from the quark-gluon side, this is mainly the physics of
    the (effective) bag constant, the difference between perturbative
    and real vacuum. Without that there is no physics of the softest
    point. If we estimate the difference between the two vacuum energy
    densities as on the order of 400 MeV/fm$^3$ we better stop 
     calculations in the ``wrong'' vacuum at energy densities
    which are large against this value (or we take 
     the difference  into account). 
    The second reason is that transport theory looses  its advantages
     to hydrodynamics  if detailed balance is given up.
    Detailed balance 
     follows from time reversal invariance. Without it, a system out of
     equilibrium is no longer driven towards equilibrium. 
    Some arbitrary ``one-way'' prescription for a ``parton--hadron''
    transition is a massive violation of detailed balance. 
     I believe that many successes (and failures) of transport models
     have to do with (no) implementation of both directions in transition
    processes. They tend to make the chemistry of the reaction insensitive
     to the microscopic transition rates. 

   (Un-)fortunately, QCD is a  complicated theory.
    It should be treated like that.

\end{document}